\documentclass[journal]{IEEEtran}
\usepackage{amsmath,epsfig,booktabs}
\usepackage{amsbsy,subfigure}
\usepackage[ruled,vlined]{algorithm2e}
\usepackage{listings}
\usepackage{hyperref}
\usepackage{cite}
\usepackage{tikz}

\newtheorem{theorem}{Theorem}

\newtheorem{remark}[theorem]{Remark}
\newcommand{\ncc}{\newcommand}
\ncc{\bs}{\boldsymbol}

\newcommand{\G}{\bs{G}}

\newcommand{\X}{\bs{X}}

\newcommand{\A}{\bs{A}}

\newcommand{\x}{\bs{x}}

\newcommand{\W}{\bs{W}}

\newcommand{\thet}{\bs{\theta}}

\newcommand{\Om}{\bs{\Omega}}

\newcommand{\s}{\sigma^2}

\newcommand{\B}{\bs{B}}

\newcommand{\U}{\bs{U}}

\newcommand{\tr}{\mbox{tr}}

\newcommand{\I}{\bs{I}}

\newcommand{\bmu}{\bs{\mu}}
\newcommand{\bSigma}{\bs{\Sigma}}

\def\x{{\mathbf x}}

\begin{document}

\title{Classification of Big Data with Application to Imaging Genetics}

\author{Magnus O. Ulfarsson,~\IEEEmembership{Member,~IEEE,}
        Frosti Palsson,~\IEEEmembership{Student Member,~IEEE,} Jakob Sigurdsson~\IEEEmembership{Student Member,~IEEE,} Johannes R. Sveinsson,~\IEEEmembership{Senior Member,~IEEE}
\thanks{M.O.U., F.P., J.S. and J.R.S. are with the Faculty of Electrical and Computer Engineering, University of Iceland, Iceland, M.O.U. is also with deCODE genetics/Amgen, Reykjavik Iceland. This work was partly supported by the Research Fund of the University
of Iceland, the Icelandic Research Fund (130635-051), and the European Community's Seventh Framework Programme (FP7/2007–2013, grant agreement no. 602450, IMAGEMEND).  }}





\maketitle
\begin{abstract}
Big data applications, such as medical imaging and genetics, typically generate datasets that consist of few observations $\bs{n}$ on many more variables $\bs{p}$, a scenario that we denote as $\bs{p \gg n}$. Traditional data processing methods are often insufficient for extracting information out of big data. This calls for the development of new algorithms that can deal with the size, complexity, and the special structure of such datasets.  In this paper, we consider the problem of classifying $\bs{p\gg n}$ data and propose a classification method based on linear discriminant analysis (LDA).  Traditional LDA depends on the covariance estimate of the data, but when $\bs{p \gg n}$ the sample covariance estimate is singular.  The proposed method estimates the covariance by using a sparse version of noisy principal component analysis (nPCA).  The use of sparsity in this setting aims at automatically selecting variables that are relevant for classification.  In experiments, the new method is compared to state-of-the art methods for big data problems using both simulated datasets and imaging genetics datasets.
\end{abstract}

\begin{IEEEkeywords}
Image Classification, Linear Discriminant Analysis, Support Vector Machines, Noisy Principal Component Analysis, Factor Analysis, $l_0$ Vector Penalty, Imaging Genetics.
\end{IEEEkeywords}

\section{Introduction}
\IEEEPARstart{R}{ecent} technological achievements and globalization have increased data acquisition
capability in almost all corners of human activities, ranging from
scientific and engineering endeavors such as genomics, medical imaging, remote sensing,
economics and finance, and all the way to people's personal lives with the emergence of social media
through the world wide web and mobile networks. The enormous growth of data creates daunting challenges,
not only in finding out how to store and access the data, but more importantly, how to process
and make sense of it. Also, since data collection is expensive, we are somehow obliged to make good use of the data at hand, so it is obvious that for further progress, the development of efficient algorithms for processing big data is very important. 

Big data is usually considered in terms of the number of observations $n$ and the number of variables $p$ measured on each observation.  In many branches of science such as genetics and medical imaging, the number of variables is very large and is often much larger than the number of observations. This scenario is often denoted as $p \gg n$.  This makes information extraction challenging due to high variance and risk of overfitting.  In the last decade, much effort has been put into developing and analyzing algorithms for $p \gg n$ problems in statistical signal processing and related fields.

An important problem in dealing with big data is how to design and build algorithms that recognize and classify patterns.  This problem is in the realm of statistical learning \cite{Hastie09,bishop,Duda01}.  There are three types of statistical learning problems, i.e., supervised learning, unsupervised learning, and reinforcement learning.  Supervised learning, also known as classification, is based on using labeled training data for constructing a function that is able to classify new observations. On the other hand, unsupervised learning, also known as clustering, is based on using distance measures to find hidden structures in unlabeled data.  Finally, reinforcement learning \cite{Kaelbling96} operates in a real-time setting where the classifier uses a feedback mechanism to improve results.  In this paper, we are concerned with supervised learning, i.e., classification.  The development and use of classification algorithms is important in many applications areas, including remote sensing \cite{Fauvel13}, handwritten digit recognition \cite{LeCun98}, face recognition \cite{Chellappa95}, DNA expression microarrays \cite{Golub99}, and functional magnetic resonance imaging \cite{Pereira09}.

A rich source of high-dimensional data is structural magnetic resonance imaging (MRI).
MRI is a noninvasive method that can be used to visualize the living human brain.  It delivers detailed brain maps in high spatial resolution that have proven to be very useful, both in a clinical and research setting, for detecting structural features in the brain.  A typical imaging experiment acquires MRI data from two groups of people; a control group, and a group consisting of subjects having some specific condition, such as a disease. In \cite{Kloppel08}, it was shown that a support vector machine (SVM) classifier can successfully distinguish subjects with mild clinically probable Alzheimer's disease from age/sex matched controls in 89$\%$ of cases; \cite{Gould14} used an SVM classifier to investigate how well certain subtypes of schizophrenia patients can be discriminated from healthy controls; \cite{Kawasaki07} developed a classification method based on discriminant analysis for differentiating between schizophrenia patients and controls; \cite{Niewenhuis12} classified 71.4$\%$ of schizophrenia patients correctly from healthy controls in a large sample using an SVM; \cite{Schnack14} proposed to use an SVM classifier to separate bipolar patients from schizophrenia patients.

Since the completion of the draft sequence of the human genome \cite{Lander01}, there has been an explosion in the availability of genomic data.  This has been driven by the development of more efficient sequencing and genotyping technologies.  The availability of genetic data has made it possible to investigate the association of genetic variants with a disease, e.g., Alzheimer's disease \cite{Jonsson12}, or a quantitative phenotype, e.g., height \cite{Gudbjartsson08}. There are several kinds of DNA sequence variations and these include single-nucleotide polymorphisms (SNPs), which are variations at the single nucleotide level that occur commonly in the population, and copy-number variations (CNVs), which are deletions or duplications of genomic regions, ranging from a several base pairs to large fractions of chromosomes.


The concurrent availability of genomic and neuroimaging data has launched the field of imaging genetics \cite{Lindenberg06}.  The aim of imaging genetics is to view neuroimaging data as a multivariate phenotype and investigate how genetic variations affect the brain.  A voxelwise genomewide association (vGWAS) was used in \cite{Stein10} to search for genetic variants that influence brain structure; \cite{Vounou10,Vounou12} developed a sparse reduced-rank method for vGWAS.  The effect of a certain copy
number variation (CNV), that confers a risk for schizophrenia, on the brain was demonstrated in \cite{Stefansson13}.  Currently, there are large-scale imaging genetics projects in operation, e.g., \cite{Stein12,Hilbar14}.

The size of the datasets involved in imaging genetics problems is typically very large and of the $p \gg n$ type.  Here, a novel method for classifying $p \gg n$ data is developed.  The method is based on linear discriminant analysis (LDA) and uses a covariance model based on a sparse version of noisy principal component analysis (nPCA). Sparsity enables the method to automatically retain variables that are important for classification and discard the rest.  At the same time, this sparsity increases the interpretability
and also decreases computational time.  We call the new method LDA-svnPCA$_0$.

The remainder of the paper is organized as follows.  The classification problem in general is reviewed in Section II.  Section III presents LDA in the $p \gg n$ settings, and discusses the use of the nPCA covariance estimate for LDA.  In Section IV, the proposed LDA-svnPCA$_0$ method is detailed.  Section V describes the structural MRI data used in this paper and the associated feature extraction and selection methods used in the experiments. Then, in Section VI, experimental results are presented.  Finally, in Section VII the conclusion is drawn.

\subsection{Notation}
Vectors and matrices are denoted using bold faced symbols. Row vector number $i$ of a matrix $\X$ is denoted by $\x_i^T$, column vector $j$ is denoted by $\x_{(j)}$; the $i$th element of a vector $\x$ is denoted as $x_i$; $\mbox{sign}(\x)$ is a vector of the same size of $\x$ with its $i$th element equal to the sign of $x_i$.  The hinge loss function is defined as $|x|_+ = x$ if $x \geq 0$ and $0$ otherwise; $\x \sim N(\bmu,\bs{\Sigma})$ means that the random vector $\x$ is Gaussian distributed with mean $\bmu$ and covariance $\bs{\Sigma}$; $I(x \geq h)$ is the indicator function that is equal to one if the inequality is true and zero otherwise; $E[x]$ denotes the expectation of a random variable $x$; $\bs{I}_p$ is the $p \times p$ identity matrix.



\section{Classification}
The general setup of a classification problem is that there is an object that belongs to one of $K$ classes $C_1,...,C_K$ and the aim is to determine which class it belongs to.  This object could for example be an individual that has one of $K$ different genetic variants.  Associated with the object is a measurement of a feature vector $\x$ and behind each feature vector is a true unknown discrete valued function $c(\x)$ where $c(\x)=k$ if and only if $\x \in C_k$.  A classifier is an estimator $\hat{c}(\x)$ of this discrete valued function.

The first step in measuring the performance of a classifier is to define a loss function that quantifies the cost associated with misclassification.  A commonly used loss function is the 0-1 loss given by
\begin{eqnarray}
L(k,l)=\left \{ \begin{array}{c c} 1, & l \neq k \\ 0, & l=k. \end{array} \right. \nonumber
\end{eqnarray}
Associated with this loss function is a risk function which is the expected loss of a classifier $\hat{c}$ when the true class is $k$, i.e.,
\begin{eqnarray}
R(k,\hat{c})&=&E[L(c,\hat{c}) | c=k] \nonumber \\
&=& \sum_{l=1}^K L(k,l) Pr(\hat{c} = l | c = k) \nonumber \\
&=&  Pr(\hat{c} \neq k | c=k). \nonumber
\end{eqnarray}
The risk is simply equal to the probability of a misclassification.
The performance of a classifier is measured by the total risk, or the so-called Bayes risk given by
\begin{eqnarray}
R_{\pi}(\hat{c}) &=& E[R(k,\hat{c})] \nonumber \\
&=& \sum_{k=1}^K Pr(\hat{c} \neq k | c=k) \pi_k, \nonumber
\end{eqnarray}
where $\pi_k$ is the prior probability of the $k$th class.  It can be shown \cite{bishop} that a classifier that minimizes the Bayes risk
is given by the Bayes classifier
\begin{eqnarray}
\hat{c}(\x)= \operatornamewithlimits{argmax}_l p(l|\x,\thet), \label{classifier}
\end{eqnarray}
that assigns a new sample to the most probable class according to the posterior class probability density function (pdf)
\begin{eqnarray}
p(l| \x, \thet)= \frac{\pi_l p_l(\x,\thet)}{\sum_{k=1}^K \pi_k p_k(\x,\thet)}, \nonumber
\end{eqnarray}
where $p_l(\x; \thet)$ is the pdf of class $l$, and $\thet$ is a parameter vector.  In the last few decades, many different classifiers have been developed and shown to be useful for a wide variety of applications.
These classifiers, include the $k$-nearest neighbors, neural networks, random forests \cite{Breiman01}, penalized logistic regression \cite{Zhu04}, support vector machines \cite{Vapnik96,Scholkopf01}, etc.
\subsection{Support Vector Machines}
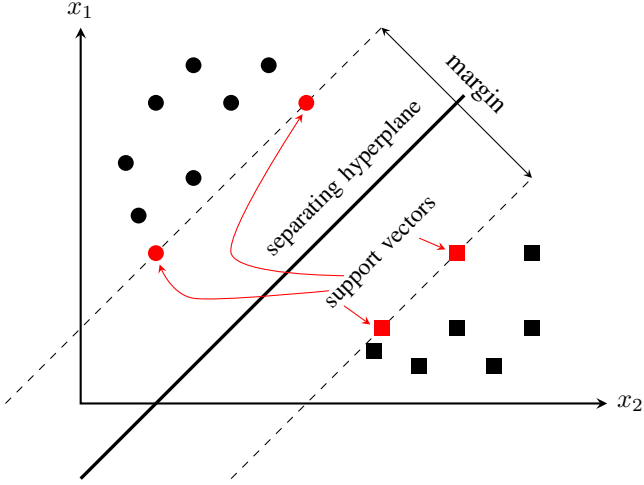
\begin{figure}%
\makeatletter
\def\pgf@plot@curveto@handler@finish{%
  \ifpgf@plot@started%
    \pgfpathcurvebetweentimecontinue{0}{0.995}{\pgf@plot@curveto@first}{\pgf@plot@curveto@first@support}{\pgf@plot@curveto@second}{\pgf@plot@curveto@second}%
  \fi%
}
\makeatother

\newcommand{\myRect}[3]{
\fill[#3]   (#1-0.1,#2-0.1)   rectangle(#1+0.1,#2+0.1);
}

\begin{tikzpicture}[>=stealth,auto] 
  \draw [<->,thick] (0,5) node (yaxis) [above] {$x_1$} |- (7,0) node (xaxis) [right] {$x_2$};
  \draw[line width=1.25pt] (0,-1) -- (5.1,4.1);
  \draw[dashed] (-1,0) -- (4,5);
  \draw[dashed] (2,-1) -- (6,3);
  \draw (3.5,3) node[rotate=45,font=\small]  {separating hyperplane};
  \draw (4,2) node[rotate=45,font=\small]  {support vectors};
  \draw[<->] (4,5) -- (6,3);
  \draw (5.25,4.25) node[rotate=-45] {margin};

  \fill[red]   (1,2)   circle (3pt);
  \fill[red]   (3,4)   circle (3pt);
  \fill[black] (1,4)     circle (3pt);
  \fill[black] (2.5,4.5)    circle (3pt);
  \fill[black] (2,4)   circle (3pt);
  \fill[black] (0.77, 2.5) circle (3pt);
  \fill[black] (1.5,3)     circle (3pt);
  \fill[black] (1.5,4.5)   circle (3pt);
  \fill[black] (0.6,3.2)   circle (3pt);

	\myRect{4}{1}{red}
	\myRect{5}{2}{red}
	\myRect{6}{2}{black}
	\myRect{4.5}{.5}{black}
	\myRect{3.9}{0.7}{black}
	\myRect{5}{1}{black}
	\myRect{6}{1}{black}
	\myRect{5.5}{0.5}{black}

\draw [->,red] plot [smooth] coordinates {(3.3,1.5) (1.5,1.4)  (1.05,2-0.15)};
\draw [->,red] plot [smooth] coordinates {(3.5,1.7) (2,2)  (3-0.05,4-0.15)};
\draw [->,red] plot [smooth] coordinates {(3.5,1.3)   (4-0.15,1.05)};
\draw [->,red] plot [smooth] coordinates {(4.5,2.2)   (5-0.15,2+0.05)};

\end{tikzpicture}
\caption{The linear SVM illustrated.}%
\label{fig:svm}%
\end{figure}

Among the most successful classification methods are the support vector machines (SVMs).  There are two kinds of SVMs, non-linear and linear, and for $p \gg n$ the linear SVM has been found to work as well as the non-linear versions \cite{Hastie09}.  The two-class linear SVM is given by
\begin{eqnarray}
\hat{c}(\x) = \mbox{sign}(f(\x)),  \nonumber
\end{eqnarray}
where $f(\x)=\bs{\beta}_0 + \bs{\beta}^T \x$.  The main idea behind linear SVM is to find a separating hyperplane, i.e., determine $\bs{\beta}_0$ and $\bs{\beta}$, that maximize the margin that separates the closest observations from either class.  Fig. \ref{fig:svm} shows a diagram illustrating the principle of the linear SVM.

Given training data $(\x_i, c_i),~i=1,...,n$, where $c_i \in \{-1,1\}$ is a label, the SVM problem can be formulated as a penalized regression problem \cite{Wahba00}
\begin{eqnarray}
(\hat{\bs{\beta}}_0,\hat{\bs{\beta}})=\operatornamewithlimits{argmin}_{\bs{\beta}_0,\bs{\beta}} \sum_{i=1}^n [1-c_i f(\x_i)]_+ + \frac{1}{2C} \| \bs{\beta} \|^2. \label{SVM}
\end{eqnarray}

The first term is the so-called hinge loss which is a convex function that upper bounds the 0-1 loss function. The second term is a regularization term that encourages smoothness.
Provided a reasonable choice of the tuning parameter $C$ and if the training data is separable, this problem delivers a decision boundary $\{f(\x)=0\}$ that maximizes the margin between the two classes.  It turns out that for $p \gg n$ problems the choice of $C=\infty$ (no regularization) separates the data \cite{Hastie09}.  However, some amount of regularization is often preferable and this means that the tuning parameter $C$ has to be selected.  The papers \cite{Hastie04,Solo05} discuss automatic ways for choosing this parameter.

The formulation (\ref{SVM}) allows for various extensions of the method, e.g., \cite{Chapelle07,Lee01,Zhou10} investigate smooth alternatives to the hinge loss function while sparse extensions of SVM are proposed in \cite{Zhu04,Fung04,Mangasarian06,Wang06,Zhang06,Liu07,Liu07b}.



\section{Linear Discriminant Analysis}
LDA is probably one of the most commonly used classification techniques.  Similarly to the linear SVM, it aims to find a linear decision boundary that separates the classes of interest.  LDA assumes that the class probabilities are multivariate normal distributions with a common covariance matrix $\bs{\Sigma}$ of size $p \times p$ and centroids $\bmu_k, k=1,...,K$, i.e., the pdf of the $k$th class is given by $p_k(\x,\thet) \sim N(\bmu_k,\bs{\Sigma})$.

In this Gaussian framework, it is easier to work with the so-called discriminant function $\delta_k(\x)=\log(p_k(\x,\thet)\pi_k)$ which is given by (ignoring irrelevant terms)
\begin{eqnarray}
\delta_k(\x) = \x^T \bSigma^{-1} \bmu_k - \frac{1}{2} \bmu_k^T \bSigma^{-1} \bmu_k + \log \pi_k, \label{disc}
\end{eqnarray}
and then assign the sample $\x$ to a class according to the decision rule (equivalent to (\ref{classifier}))
\begin{eqnarray}
\hat{c}(\x)=\operatornamewithlimits{argmax}_k \delta_k(\x).
\end{eqnarray}
In practise, the parameters $\bs{\Sigma}$ and $\bmu_k, k=1,...,K$ need to be estimated from the data.  Given training data $(\x_{i},c_i),~i=1,...,n$ where $n>p$ one traditionally uses the following estimates:
\begin{eqnarray}
\hat{\bmu}_k &=& \frac{1}{n_k}\sum_{i \in C_k} \x_{i}, \quad k=1,...,K \label{eq1}, \\
\hat{\bs{\Sigma}} &=& \frac{1}{n-K} \sum_{k=1}^K \sum_{i \in C_k} (\x_{i}-\hat{\bmu}_k)(\x_{i}-\hat{\bmu}_k)^T, \label{eq2} \\
\pi_k &=& \frac{n_k}{n}, \quad k=1,...,K, \label{eq3}
\end{eqnarray}
where $C_k=\{i: c_i=k\}$, and $n_k=|C_k|$. We note that the above is not the only way for motivating LDA.  It can also be motivated from optimal scoring, and Fisher's discriminant analysis \cite{Mardia79}.

\subsection{The Big Data Case: $p \gg n$}
In the big data case, where one has many more variables than samples ($p \gg n$), the covariance estimate (\ref{eq2}) is singular and therefore the LDA classifier cannot be constructed.  In this case, researchers have suggested to
constrain the estimate to be positive definite.  In \cite{Tibshirani02,Bickel04,Pang09}, the covariance was constrained to be diagonal, i.e., $\hat{\bs{\Sigma}}_d=\mbox{diag}(\hat{\sigma}_1^2,...,\hat{\sigma}_p^2)$, where $\hat{\sigma}_j^2$ is the $j$th diagonal element of (\ref{eq2}).  An alternative approach for avoiding the singularity problem was presented in \cite{Guo07} where the regularized covariance estimate
\begin{eqnarray}
\tilde{\bSigma}=\alpha \hat{\bSigma}+(1-\alpha)\bs{I}_p \label{cm1}
\end{eqnarray}
is used for some $\alpha>0$.

Another problem with LDA in high dimensions is that the classifier is a linear combination of all the $p$ variables which hinders interpretability.  A solution to this problem
is to select a small subset of the variables, while conserving the discriminative power of the classifier.  The paper \cite{Tibshirani03} proposed the nearest shrunken centroid (NSC) method that is based on traditional LDA using a diagonal covariance matrix.  The idea is to shrink the class centroids toward the overall mean.  In detail, the $k$th NSC discriminant function is given by
\begin{eqnarray}
\delta_k(\x) = \sum_{j=1}^p \frac{x_j \tilde{\mu}_{kj} - \frac{1}{2} \tilde{\mu}_{kj}^2}{\hat{\sigma}^2_j} + \log \pi_k, \nonumber
\end{eqnarray}
where the shrunken centroids $\tilde{\mu}_{kj}$ are given by
\begin{eqnarray}
\tilde{\mu}_{kj}&=&\hat{\mu}_j+(\hat{\sigma}_j+\gamma)\tilde{d}_{kj}, \nonumber \\
\tilde{d}_{kj}&=&\max(|d_{kj}|-h,0)\mbox{sign}(d_{kj}), \nonumber \\
d_{kj}&=&\frac{\hat{\mu}_{kj}-\hat{\mu}_j}{m_k(\hat{\sigma}_j+\gamma)}, \nonumber
\end{eqnarray}
where $\hat{\mu}_{kj}$ is the $k$th centroid, $\hat{\mu}_j$ is the overall mean, $m_k=\sqrt{\frac{1}{n_k}-\frac{1}{n}}$, and $\gamma$ is a small positive constant.
By increasing $h$, the method zeroes out some of the $\tilde{d}_{kj}$ variables, which in turn means that the corresponding variables $x_j$ do not contribute to the class prediction.  The tuning parameter $h$ has to be selected, \cite{Tibshirani03} suggests to use $L$-fold cross-validation.

In \cite{Guo07}, the shrunken centroid idea was combined with using the regularized covariance estimate (\ref{cm1}) in the LDA framework.
A covariance model based on the noisy principal component analysis (nPCA) (see next subsection) was used in \cite{Li12} and a sparse optimization framework was used to select variables. The LDA was formulated as an $l_1$ penalized optimal scoring problem in \cite{Clemmensen11} to achieve automatic variable selection; and in \cite{Witten12} LDA was formulated as an $l_1$ penalised Fisher's linear discriminant problem with the same objective. The paper \cite{Wu15} provides a discussion and compares the formulation of sparse optimal scoring and sparse Fisher's discriminant analysis.

In the following, the singularity issue of the covariance matrix is solved by using an estimate that is positive definite and relates to nPCA.       

\subsection{Noisy Principal Component Analysis}
The nPCA model is a multivariate model of the following
form
\begin{eqnarray}
\x = \bmu + \G \bs{u} + \bs{\epsilon}, \label{npca_model}
\end{eqnarray}
where $\bmu$ is the mean vector, $\G$ is a $p \times r$ matrix, $p \gg r$,  $\bs{u} \sim N(\bs{0},\bs{I}_r)$ contains $r$ noisy principal components, $\bs{\epsilon} \sim N(\bs{0},\s \bs{I}_p$) is noise, and $\bs{u}$ and $\bs{\epsilon}$ are independent.  It can be shown that $\x \sim N(\bmu,\Om)$ where $\Om=\G \G^T + \s \bs{I}_p$ is the nPCA model of the covariance.

The nPCA model has been found to be useful in signal processing applications such as array signal
processing \cite{Wax85}, and medical imaging applications \cite{Ulfarsson08b}.  The model was originally developed by \cite{Lawley53} but later popularized under the name
probabilistic PCA \cite{Tipping99}, however, we prefer to call it nPCA.

Given a sample $\x_i,i=1,...,n$, the unknown model parameters
$\bmu,\boldsymbol{G},\sigma^2$ are estimated using
the maximum likelihood principle, yielding \cite{Lawley53}
\begin{eqnarray}
\hat{\bmu}&=& \frac{1}{n} \sum_{i=1}^n \x_i, \nonumber \\
\hat{\boldsymbol{G}} &=& \boldsymbol{P}_r(\boldsymbol{L}_r -
\hat{\sigma}^2 \boldsymbol{I}_r)^{1/2} \boldsymbol{R}, \label{npca1} \\
\hat{\sigma}^2 &=& \frac{1}{M-r} \sum_{j=r+1}^M l_j, \label{npca2}
\end{eqnarray}
where $\boldsymbol{L}_r=\mbox{diag}(l_1,...,l_r)$ contains the $r$
largest eigenvalues of the data covariance matrix
$\boldsymbol{S}_{x}=\frac{1}{n}\sum_{i=1}^n(\x_i-\hat{\bmu})(\x_i-\hat{\bmu})^T$, $\boldsymbol{R}$ is an arbitrary
$r \times r$ orthogonal rotation matrix, and $\boldsymbol{P}_r$
contains the $r$ first eigenvectors of
$\boldsymbol{S}_{x}$ in its columns. In practice, $\bs{R}$ can be set equal to the identity matrix.  Alternatively, the nPCA model parameters can be computed using an expectation-maximization (EM) algorithm \cite{Tipping99}.

The nPCA model can be extended in various ways.  In \cite{Ulfarsson11}, a method called sparse variable nPCA (svnPCA) was proposed that is based on performing nPCA while automatically discarding irrelevant variables from the analysis.

A closely related model to the nPCA model is the factor analysis (FA) model \cite{Bartholomew87}.  The difference is that the noise term in (\ref{npca_model}) is $\bs{\epsilon} \sim N(\bs{0},\bs{\Psi})$ where $\bs{\Psi}=\mbox{diag}(\s_1,...,\s_p)$.  In this case, the observed data is also Gaussian distributed but with a different covariance matrix $\Om=\G \G^T + \bs{\Psi}$.  Unlike nPCA, there are no closed form solutions for $\G$ and $\bs{\Psi}$, so an iterative algorithm is needed to estimate them.

\subsection{Linear Discriminant Analysis using Noisy Principal Component Analysis}
Our motivation for introducing nPCA is to use it to estimate the covariance $\Om$ of the observed data.  The benefit of using this method is the large difference in the number of covariance parameters that needs to be estimated.  In the case of nPCA, the number of covariance parameters is $pr+1-r(r-1)/2$. 
This can be contrasted with the number of parameters in the full covariance model which is $p(p+1)/2$, which is much larger if $r$ is a relatively small number.  A discussion of the use of nPCA, and the closely related FA, as a covariance model can be found in \cite{Fan11,Tipping99}.  

In the following, we propose to use the nPCA covariance model $\Om$ in the LDA model.  This involves exchanging $\bs{\Sigma}^{-1}$ for $\Om^{-1}$ in (\ref{disc}).  In our applications, the variable dimension $p$ is very high and therefore the construction of $\Om^{-1}$ is unfeasible.  Instead, we use the matrix inversion lemma
\begin{eqnarray}
\Om^{-1}&=& \frac{1}{\s}\bs{I}_p - \frac{1}{\s} \G \W^{-1} \G^T, \nonumber \\
\W&=& \s \bs{I}_r + \G^T \G, \nonumber
\end{eqnarray}
and note that to construct the discriminant function it is only necessary to construct $\Om^{-1}\bmu_k$, which is much cheaper than having to explicitly form the covariance.


 \section{Linear Discriminant Analysis using Sparse Noisy Principal Component Analysis}
 As stated above, the lack of interpretability is a problem in the $p \gg n$ scenario. In this paper, we propose a method to automatically drop out variables that do not contribute to the class prediction.  To begin reformulating the LDA problem for better
 interpretation, we write the model for class $k$ in terms of the distance of the $k$th centroid from the mean vector, i.e., $\bs{d}_k= \hat{\bmu}_k-\hat{\bmu}$.  Then the discriminative function for class $k$ can be written as
 \begin{eqnarray}
 \delta_k(\x)=\tilde{\x}^T \Om^{-1} \bs{d}_k-\frac{1}{2} \bmu_k^T \Om^{-1} \bs{d}_k + \log  \pi_k, \nonumber
 \end{eqnarray}
 where $\tilde{\x}=\x-\hat{\bmu}$.  Now we want to identify variables $\tilde{x}_j$ that can be discarded from the analysis while maintaining the class discrimination.
 First note that if the $j$th element of $\Om^{-1} \bs{d}_{k}$ is zero then the $j$th variable can be discarded from the classifier.  Now note that this element can be written as
 \begin{eqnarray}
 [\Om^{-1} \bs{d}_k]_j = \frac{d_{kj}}{\s}-\frac{1}{\s}\bs{g}_j^T \W^{-1} \G^T \bs{d}_{k}, \nonumber
 \end{eqnarray}
 where $\bs{g}_j^T$ is the $j$th column of $\G$.  From this formula, we see that the variable $j$ does not play a part in the $k$th discriminative function if $d_{kj}=0$ and $\bs{g}_j=\bs{0}$.  So in general, we can discard the $j$th variable if all the elements
 in the vector
\begin{eqnarray}
\bs{a}_j = [d_{1j}, d_{2j},..., d_{Kj}, \bs{g}^T_j]^T \nonumber
\end{eqnarray}
 are zero.

 The method we propose, which we call LDA-svnPCA$_0$, is based on maximizing the discriminant functions, while automatically discarding variables that do not contribute to the class prediction, by encouraging the $\bs{a}_j$ vectors to be zero.  Given training data $(\x_i,c_i),~i=1,...,n$, we propose to estimate the parameters $\thet=\{\G,\s,\bs{d}_1,...,\bs{d}_K\}$ of the classifier by solving the following optimization problem
 \begin{eqnarray}
 \hat{\thet} = \operatornamewithlimits{argmax}_{\thet} J_{\thet}(\mathcal{X}), \label{opt}
 \end{eqnarray}
 where $\mathcal{X}=\{\x_1,...,\x_n\}$ is the observed data, and $J_{\thet}(\mathcal{X})$ is a penalized discriminant function given by
 \begin{eqnarray}
 J_{\thet}(\mathcal{X})= \frac{1}{n}\sum_{k=1}^K \sum_{\x_i \in C_k}  \delta_k(\x_i)-\frac{h}{2} \sum_{j=1}^p |||\bs{a}_{j}|||_0. \label{cost}
 \end{eqnarray}
  This cost function aims to maximize the discriminant function, $\delta_k(\x_i)$, for each class while using the vector $l_0$ penalty, $|||\bs{a}_{j}|||_0$, to enforce some of the $\bs{a}_{j}=[d_{1j},...,d_{Kj}, \bs{g}_j^T]^T$ to be zero.  The vector $l_0$ penalty works in the following way:  $|||\bs{a}_{j}|||_0 = I(\bs{a}_{j} \neq \bs{0})=1$ if $\|\bs{a}_{j}\| \neq 0$, and otherwise it equals 0.  In other words, when $\bs{a}_j \neq \bs{0}$, then $\frac{h}{2}$ is subtracted from the criterion, and when $\bs{a}_j=\bs{0}$ nothing gets subtracted.  The vector $l_0$ penalty has previously been used in other context, e.g., for sparse principal component analysis \cite{Ulfarsson11}; sparse independent component analysis \cite{Palsson14}; multivariate regression \cite{Seneviratne12}; hyperspectral denoising \cite{Palsson14IGARSS}; magnetoencephalography (MEG) \cite{Luessi14,Cassidy15}.  It either totally removes or keeps the vector $\bs{a}_j$, and it should not be confused with the scalar $l_0$ penalty $\|\bs{a}_j\|_0=\sum_i I( a_{ij} \neq 0)$ which only removes or keeps individual elements of $\bs{a}_j$.

 The optimization problem (\ref{opt}) does not have a closed form solution so we need to resort to an iterative algorithm.  Fortunately, there is an efficient EM algorithm \cite{Dempster77} that can solve this problem.

\subsection{EM Algorithm}
The EM algorithm is an iterative algorithm that is often efficient at optimizing likelihood functions.  Typically, one is interested in estimating a parameter vector $\thet$ from a likelihood $p_{\thet}(\x)$.  The EM algorithm is primarily useful when the model behind $p_{\thet}(\x)$ depends on a missing data vector $\bs{u}$, that observed would make the estimation of $\thet$ easy. The usefulness of the algorithm is demonstrated with numerous examples in \cite{Moon96}.

The algorithm is centered around the EM functional which minorizes $p_{\thet}(\x)$. The EM functional is constructed in the expectation (E) step of the algorithm.  Then, in the maximization (M) step, the EM functional is maximized  w.r.t. $\thet$.  This process is repeated until convergence.  Issues relating to the convergence of the algorithm are discussed in \cite{Wu83} and \cite{Lange95}.


\subsection{EM Algorithm for the Proposed Method}
As stated in the previous section, the EM algorithm consists of performing the E- and the M-step of the algorithm in an iterative manner.  Below, we develop those steps for the
LDA-svnPCA$_0$ algorithm.  Note that this is an iterative algorithm and we denote the current iterate of a parameter with a subscript 0, and the next iterate with a subscript 1.

The E-step consists of constructing the EM functional.  First, it is necessary to construct the complete penalized discriminant function which is the discriminant function for the case where the (missing) data $\mathcal{U}=\{{\bs{u}}_1,...,{\bs{u}}_n \}$ is observed.  The complete penalized discriminant function is (ignoring irrelevant terms)
\begin{eqnarray}
J_{\thet}(\mathcal{X},\mathcal{U})&=& -\frac{p}{2} \log \s - \frac{1}{n} \sum_{k=1}^K \sum_{i \in C_k} \frac{\|\tilde{\x}_i - \bs{d}_k-\G \bs{u}_i\|^2}{2\s} \nonumber \\
&-& \frac{h}{2} \sum_{j=1}^p |||\bs{a}_j|||_0.  \nonumber
\end{eqnarray}
The EM functional $\mbox{EM}(\thet_0,\thet)=E_{\thet_0}[ J_{\thet}(\mathcal{X},\mathcal{U})|\mathcal{X}]$ is the expected value of the complete penalized discriminate function w.r.t.
\begin{eqnarray}
p(\bs{u}_i|\tilde{\bs{x}}_i,\G_0,\sigma_0^2)=N(\W_0^{-1}\G_0^{T}(\tilde{\bs{x}}_i-\bs{d}_k),\sigma_0^2 \W_0^{-1}), \nonumber
\end{eqnarray}
 and leads to
\begin{eqnarray}
\mbox{EM}(\thet_0,\thet) &=& -\frac{p}{2} \log \s - \frac{1}{n}\sum_{k=1}^K \sum_{i \in C_k} \frac{\|\tilde{\x}_i - \bs{d}_k-\G \bs{u}_{i0} \|^2}{2\s} \nonumber \\
&-& \frac{\s_0 \tr\left( \G \W_0^{-1} \G^T \right)}{2\s} - \frac{h}{2} \sum_{j=1}^p |||\bs{a}_j |||_0,  \label{EM}
\end{eqnarray}
where $\W_0=\G_0^T \G_0 + \s_0 \bs{I}_r$, and $\bs{u}_{i0}=\W_0^{-1}\G_0^T(\tilde{\x}_i-\bs{d}_{k0})$.

In the M-step of the EM algorithm we optimize the EM functional to obtain the update for the model parameter.  The update formulas depend on the following quantities
\begin{eqnarray}
\bs{U}&=&[\bs{u}_{i0}^T], \nonumber \\
\bs{A}_0&=& \s_0 \bs{W}_0^{-1} + \frac{1}{n} \bs{U}^T \bs{U}, \nonumber \\
\bs{B}_0 &=& \frac{1}{n} \tilde{\X}^T \U. \nonumber
\end{eqnarray}
The model parameter updates also depend on the following thresholding parameters
\begin{eqnarray}
\tau_j^2 = \bs{b}_{j0}^T \A_0^{-1} \bs{b}_{j0} + \sum_{k=1}^K \frac{n_k}{n}(\hat{\mu}_{kj}-\hat{\mu}_j)^2, \quad j=1,...,p \nonumber
\end{eqnarray}
that are used to determine which variables are dropped from the model.  The model parameter updates are given by (see the appendix for a derivation)
\begin{eqnarray}
d_{kj1}&=& (\hat{\mu}_{kj}-\hat{\mu}_j)I(\tau_j^2 \geq h \s_{0}), \quad j=1,...,p, \nonumber \\
\bs{g}_{j1} &=& \bs{A}_0^{-1} \bs{b}_{j0} I(\tau_j^2 \geq h \s_{0} ), \quad j=1,...,p, \nonumber \\
\sigma^2_1&=& \frac{1}{p} \sum_{j  \in \mathcal{I}} \left( S_{xjj}- \bs{b}_{j0}^T\A_0^{-1} \bs{b}_{j0} \right) + \frac{1}{p}\sum_{j \in \mathcal{I}^c} S_{xjj}, \nonumber
\end{eqnarray}
where $S_{xjj}$ is the $j$th diagonal element of
\begin{eqnarray}
\bs{S}_{x}=\frac{1}{n} \sum_{k=1}^K \sum_{i \in C_k} (\tilde{\x}_i-\bs{d}_{k1})(\tilde{\x}_i-\bs{d}_{k1})^T, \nonumber
\end{eqnarray}
and
$\mathcal{I}=\{j: \tau_j^2 \geq h \sigma^2_0 \}$, and $\mathcal{I}^c=\{j: \tau_j^2 < h \sigma^2_0 \}$.
For convenience the algorithm is listed in Algorithm 1.
\begin{remark}
Note that due to the dependency of the $d_{kj}$ and the $\bs{g}_j$ updates with $\sigma^2_0$, the update equations need to be iterated to maximize the M-step.  However, in our experiments we found one iteration to be sufficient.
\end{remark}
\begin{remark}
The EM algorithm depends on two tuning parameters $r$ and $h$.  In this paper, cross-validation is used to select them. Refer to Section VI B for details.
\end{remark}
\begin{remark}
Due to the non-smoothness of the vector $l_0$ penalty the convergence theory in \cite{Wu83} and \cite{Lange95} does not apply.  However, the convergence result in \cite{Ulfarsson11} does apply for the proposed EM algorithm.
\end{remark}
\begin{remark}
The computational complexity for an iteration of Algorithm 1 is only $O(prn)$.  This does not account for possible savings due to sparsity, i.e., if there are only $p_h<p$ non-zero columns of $\bs{G}_0$ then the complexity becomes $O(p_h r n)$.  There are no extra memory constraints apart from storing the data and the iterates.  In fact, if storing the data matrix is a problem due to its size it is easy to construct a sequential version of this algorithm, i.e., process one sample (row of $\tilde{\bs{X}}$) at a time.
\end{remark}

\begin{algorithm}
\DontPrintSemicolon \KwIn{Data matrix $\tilde{\X}$, $C_1$,...,$C_K$, $r$, and $h$}
\textbf{Initialization}: $\G_0$, $\hat{\bmu}$, $\bs{d}_{10}$,...,$\bs{d}_{K0}$, and $\s_0$ \; \While{(Not
converged)}{ $\bs{W}_0 = \G_0^T \G_0 + \s_0 \bs{I}_r$\; \For{$k=1,...,K$}{\For{$ i \in C_k$}{$\bs{u}_{i0}=\W_0^{-1} \G_0^T(\tilde{\x}_i-\bs{d}_{k0})$} }  $\U =
[\bs{u}_{i0}^T]$\;  $\bs{A}_0 = \s_0\bs{W}_0^{-1} +
\frac{1}{n}\U^T\U$\; $\bs{B}_0 = \frac{1}{n} \tilde{\X}^T \U$\; \For{$j=1,...,p$}{$\tau_j^2=\bs{b}^T_{j0}\A_0^{-1} \bs{b}^T_{j0}+\sum_{k=1}^K(\hat{\mu}_{kj}-\hat{\mu}_j)^2$\;
$d_{kj1}=(\mu_{kj}-\hat{\mu}_j)I(\tau_j^2 \geq h \s_{j0})$ \; $\bs{g}_{j1}=\A_{0}^{-1} \bs{b}_{j0} I(\tau_j^2 \geq h \s_{j0})$\; $S_{xjj}=\frac{1}{n} \sum_{k=1}^K \sum_{i \in C_k} \| \tilde{\x}_i-\bs{d}_{k1}\|^2$\;} $\G_{1} = [\bs{g}_{j1}^T]$\; $\bs{d}_{k1}=[d_{kj1}]$\; $\mathcal{I}=\{j: \tau_j^2 \geq h \sigma_0^2 \}$\; $\mathcal{I}^c=\{j: \tau_j^2<h \sigma_0^2  \}$ \;
$\sigma^2_1= \frac{1}{p} \sum_{j  \in \mathcal{I}} \left( S_{xjj}- \bs{b}_{j0}^T\A_0^{-1} \bs{b}_{j0} \right) + \frac{1}{p}\sum_{j \in \mathcal{I}^c} S_{xjj}$\;} 
\KwOut{$\hat{\G}$, $\hat{\bs{d}}_1,...,\hat{\bs{d}}_K$, and  $\hat{\sigma}^2$.} \caption{The LDA-svnPCA$_0$ algorithm}
\label{alg:mine}
\end{algorithm}

\subsection{Special Case: Two classes and $r=0$}
When $r=0$ then $\G=\bs{0}$ and the LDA-svnPCA$_0$ method reduces to a diagonal covariance LDA with automatic selection of variables.  Assuming we have $n_1$ samples from class 1 and $n_2$ samples from class 2, then the variables included in the classifier
satisfy the condition $\tau_j^2 \geq h \s_{0}$ where
\begin{eqnarray}
\tau_j^2 = \frac{n_1}{n}(\hat{\mu}_{1j}-\hat{\mu}_j)^2 + \frac{n_2}{n}(\hat{\mu}_{2j}-\hat{\mu}_j)^2, \nonumber
\end{eqnarray}
and now since, $n=n_1+n_2$ and $\hat{\mu}_j=\frac{n_1}{n}\hat{\mu}_{1j}+\frac{n_2}{n}\hat{\mu}_{2j}$, we can write
\begin{eqnarray}
\tau_j^2 = \frac{(\hat{\mu}_{1j}-\hat{\mu}_{2j})^2}{\left( \frac{1}{n_1}+\frac{1}{n_2} \right)}. \nonumber
\end{eqnarray}
Therefore, it can be seen that the variable $j$ is selected based on whether it satisfies the following threshold
\begin{eqnarray}
|\mathcal{T}_j| = \frac{|\hat{\mu}_{1j}-\hat{\mu}_{2j}|}{\hat{\sigma} \sqrt{ \frac{1}{n_1}+\frac{1}{n_2}}} \geq h. \nonumber
\end{eqnarray}
This is the classical two-sample t-test \cite{Snedecor89}.  So, in this special case, the LDA-svnPCA$_0$ algorithm consists of first selecting variables that pass a threshold defined by the two-sample t-test, and then performing diagonal covariance LDA on the retained variables.

\section{Structural MRI Datasets}
The MRI data used in this article were collected by deCODE Genetics\footnote{deCODE Genetics/Amgen, Reykjavik, Iceland} in a study on how rare CNVs confering high risk of schizophrenia and/or other neurodevelopmental disorders affect the physiological function for an otherwise healthy brain.  The MRI data were generated for healthy controls carrying neuropsychiatric CNVs, controls carrying CNVs not known to be associated with psychiatric disorders (Other CNVs), and controls without large CNVs (NoCNV) (See \cite{Stefansson13} for more detailed description of the recruitment and phenotyping).  The MRI was collected with an 1.5 T whole body Philips Achieva scanner.  The scans were performed with a sagittal 3D fast $T_1$-weighted gradient echo sequence (TR=8.6 ms, TE=4.0 ms, flip angle=8 degrees, slice thickness=1.2 mm, field of view =240 $\times$ 240 mm).
\subsection{Segmentation}
Each of the $n$ $p$-dimensional $T_1$ weighted (vectorized) MRIs were tissue segmented into white matter, and gray matter images using the VBM8 software (\url{http://dbm.neuro.uni-jena.de}) which is integrated into the SPM8 software (Wellcome Department of Cognitive
Neurology, Institute of Neurology, London, UK, \url{http://www.fil.ion.ucl.ac.uk/spm}) implemented in MATLAB R2014 (Mathworks Inc.,
Sherborn, MA, USA).  Note that the segmentation does not change the dimensionality of the resulting images, e.g., a white matter segmented image is still $p$-dimensional.  After the segmentation, a spatial normalization step was performed which uses the DARTEL algorithm \cite{Ashburner05} to register the tissue segments into a common coordinate system.  Good spatial normalization will tighten class clusters and reduce dimensionality.  Finally, the maps from the normalization step were modulated, i.e., intensity corrected for local volume changes during spatial normalization.  No spatial smoothing was applied. 

\subsection{Feature Selection}



Feature selection is a simple preliminary screening of variables that is often a helpful first step when dealing with $p \gg n$ data. Many effective and well known feature selection techniques have been proposed in recent years, such as the Dantzig selector \cite{Candes07}, the Lasso \cite{Tibshirani96}, adaptive Lasso \cite{Zou06b}, SCAD \cite{Fan01}, minimax concave penalties \cite{Zhang10}, and the locally weighted least squares regression methods \cite{Ruppert94}. However, these techniques are often not suitable for very high dimensional data due to the unfeasible computational cost when the number of dimensions becomes very high, and finding the optimal model is not guaranteed.


The method we have chosen to employ here is called Sure Independence Screening (SIS) \cite{Fan08}. 
A brief overview of SIS can be given using the 2-class classification problem. Let $\X$ be an $n \times p$ data matrix, where each column $\x_{(i)}$ has been standardized with zero mean and unit variance, and let $\bs{c}$ be the associated $n \times 1$ label vector. A component-wise regression is performed, yielding the $p$-vector
$$\bs{\omega} = \X^T\bs{c}.$$
This is obviously a very computationally cheap operation. The vector $\bs{\omega}$ is basically the correlation coefficients of each feature with the label vector $\bs{c}$. The next step is to sort the magnitudes of $\bs{\omega}$ in a decreasing order and then select the first $m$ values of the sorted $\bs{\omega}$ as our features. Now suppose we have $n_1$ samples from class 1 and $n_2$ samples from class 2. The component-wise regression estimate $\bs{\omega}$ can then be written as
$$\bs{\omega}=\sum_{i \in C_1} \x_{(i)} - \sum_{i \in C_2} \x_{(i)},$$
where the $j$th component of $\bs{\omega}$ is given by
$$\bs{\omega}_j = \frac{n_1 \hat{\mu}_{1j}-n_2 \hat{\mu}_{2j}}{\hat{\sigma}_j},$$
where $\hat{\mu}_{kj}$ is $j$th element of the centroid for class $k\in \{1,2\}$, and $\hat{\sigma}_j$ is the standard deviation of the $j$th feature. When the classes are of equal size, i.e., $n_1=n_2$, then each value of $\bs{\omega}$ is simply a scaled version of the two-sample t-test for the corresponding feature.

\section{Experimental results}
\def\svn{LDA-svnPCA$_0$ }
\def\FA{LDA-FA$_0$ }
\def\TEa{TE}
\def\TE2{TE$_{\mbox{opt}}$}
 In this section, the proposed method is compared to state-of-the-art classification methods for big data problems. We perform several experiments using both simulated and real data. There are two simulated datasets. In the first dataset, the covariance matrix has an independent structure, i.e., $\bs{\Sigma}=\bs{I}_p$ and the only difference between the classes is in their centroids. The second simulated dataset is generated according to the nPCA model \eqref{npca_model} and thus has a more complex covariance structure than the first simulated dataset. Again, the difference between the classes is in their centroids.

The first real dataset is the Golub dataset \cite{Golub99} which is a gene expression dataset which consists of two leukemia classes and 7129 genes. In this experiment, the objective is to correctly classify samples into two groups which represent different types of leukemia. We include this dataset in our experiments to show that our method is able to handle microarray data.

Finally, we perform two experiments using the structural MRI data, which was described in Section V. In the first MRI experiment we classify samples based on their structural MRI data into two groups. One group consists of control samples (NoCNV) while the second group has neuropsychiatric CNVs affecting cognition. The second MRI experiment involves classifying control samples into two groups which are defined by high and low polygenic risk scores (PRS) for schizophrenia. The group with high PRS has a higher risk of developing schizophrenia while the low PRS group has a lower risk.


\subsection{Comparison Methods}

The methods that are used for comparison are the shrunken centroid regularized discriminant analysis (SCRDA) method \cite{Guo07}, the nearest shrunken centroid (NCS) method \cite{tibshirani2002diagnosis,tibshirani2003class}, the penalized LDA method \cite{Witten12} and the linear SVM. All the comparison methods were implemented in the R statistical computing environment \cite{R}. 

The linear SVM has a single tuning parameter $C$ that controls the smoothness of the SVM solution.  The classification accuracies were insensitive to the choice of $C$ so the parameter was set equal to 1 in all experiments. The SVM was implemented using the R package e1071 \cite{RSVM} which is an interface for R to the popular LIBSVM software \cite{libSVM}.

The SCRDA method avoids the singularity problem of the covariance estimate $\hat{\bs{\Sigma}}$ arising when $p \gg n$ by using a regularized covariance estimate (\ref{cm1}). Another important feature of the SCRDA is that the class centroids are shrunken towards the overall centroid, and thus zeroing or eliminating features that do not contribute to the classification. The SCRDA method depends on two tuning parameters, that control the sparsity and the covariance regularization, respectively. Both tuning parameters are chosen using cross-validation (CV). The implementation used for this method can be found in the R package RDA \cite{rda}.

The NSC method is essentially a simplified version of SCRDA where the covariance is modeled as a positive definite diagonal matrix. It has a single tuning parameter which directly controls the sparsity of the classifier. As before, the optimal value of the tuning parameter was chosen based on CV. 
The implementation of NSC is given in the R package PAMR \cite{pamr}.


The penalized LDA method in \cite{Witten12} is a method that addresses the shortcomings of the classical LDA for $p\gg n$ problems, i.e., the within-class covariance matrix becomes singular and the interpretability of the classifier is low when the features are many. The method uses a diagonal estimate of the within-class covariance and applies an $\ell_1$ penalty to the discriminant vectors, which is controlled by the tuning parameter $\lambda$. We used 30 values for the sparsity tuning parameter $\lambda$ and the optimal value was chosen based on CV. This method is implemented in the R package penalizedLDA \cite{penalizedLDA}.  The methods used in the experiments are summarized in Table \ref{tab:methods}.

\begin{table}[htbp]
  \centering
  \caption{Methods used in the experiments}
    \begin{tabular}{rl}
    \addlinespace
    \toprule
    \multicolumn{2}{c}{Comparison methods} \\
    \midrule
    SVM & SVM with  a linear kernel \\
    SCRDA & Shrunken centroids regularized LDA \\
    NSC & Nearest shrunken centroids (NSC) LDA \\
		PLDA & Penalized sparse LDA using $\ell_1$ penalty \\
		\midrule
    \multicolumn{2}{c}{Proposed method} \\
		\midrule
    LDA-svnPCA0 & Penalized sparse LDA-svnPCA$_0$ based on svnPCA \\
    \bottomrule
    \end{tabular}%
  \label{tab:methods}%
\end{table}%

All the comparison methods, except the linear SVM, are based on LDA. When the number of features is very large, as is typical for $p\gg n$ problems, the interpretability of the algorithms is reduced, since the classifier output depends on all the $p$ features. All the comparison LDA based methods have the ability to drop variables that do not contribute to the classification via some sparsity penalty, i.e., $\ell_1$-penalty, shrunken centroids with covariance regularization, etc. Our method models the covariance using the nPCA model and via the $\ell_0$-penalty penalty on the variables, has the ability to drop irrelevant variables or features. This, coupled with the SIS feature selection helps to further increase the interpretability of the method and make the representation of the results more compact since only a small fraction of the original $p$ variables is retained in the trained model.

\subsection{Parameter Selection and General Experimental Procedure }
Our method depends on two tuning parameters, the sparsity parameter, $h$, and the number of nPCs, $r$, that need to be selected.  Here we describe the procedure used to select them.  There are many tuning parameter selection methods in the literature such as AIC \cite{Akaike1974}, BIC \cite{Schwartz78}, and the extended BIC \cite{Chen08}. However, the tuning parameter method most often used in classification problems is CV \cite{Stone74}.

CV is a technique to assess the prediction error of a model. There are many variants of CV methods and the one used in this paper is called $L$-fold CV.
The basic idea behind $L$-fold CV is to split the training data into $L$ roughly equally sized partitions or folds. One fold is kept for validation and the classifier is trained using the remaining $L-1$ folds, i.e., the classifier is trained on $L-1$ folds of the data and tested on the $l$th fold, where $l=1,2,...,L$. Finally, the sum of the prediction error for all the folds is used to obtain the CV estimate of the prediction error. By repeating this for each value of the tuning parameters, one can assign a CV prediction error estimate to them and thus use the CV estimate to choose the tuning parameters. 

For the simulated data and Golub data 10-fold CV is used while for the MRI data 5-fold CV is used. The Golub dataset is provided with predefined training and testing sets, while for the MRI data, we use two thirds of the available data for CV to select the tuning parameters and one third of the data for validation.

It is important to note that many parameter values can yield the same CV test error and thus the question remains how to select the tuning parameter(s) that give the optimal test results. In this paper, the four metrics given in Table \ref{tab:key} are used to assess the performance of the various methods.  The first metric is the lowest CV test error, the second, denoted
by Nonzeros, is the minimum number of features obtained for \TE2 (see below). The third metric, denoted by TE, is the minimum test error obtained for all the tuning parameter values that give rise to the same minimum CV test
error.  Finally, the fourth metric, which is denoted by \TE2,
is the minimum test error obtained by training the classifier
using the entire training set and iterating over the whole tuning
parameter space. Note that \TE2 can be regarded as an indicator of the optimal performance of the classifier, given the data and predefined tuning parameter values. The TE metric depends entirely on the method used to choose the optimal tuning parameters and there are many methods to do that. Hence, \TE2 is more informative regarding how well a given classifier actually performs.  

\begin{table}[htbp]
  \centering
  \caption{Performance metrics used in all experiments}
    \begin{tabular}{rl}
    \addlinespace
    \toprule
    Name & meaning \\
    \midrule
    CV err & The minimum CV test error \\
    Nonzeros & The min. \# of features obtained for optimal \TE2 \\
    \TEa & min test error (TE) obtained for optimal param. values \\
    \TE2 & min TE obtained over whole tuning parameter space \\
    \bottomrule
    \end{tabular}%
  \label{tab:key}%
\end{table}%

SIS feature selection is used to reduce the amount of features prior to classification using the proposed method, typically by a factor of hundred and, in some cases, by a factor of thousand. SIS is not used for the comparison methods, primarily since they are self-contained and complete methods. Also, feature selection does not work well with linear SVM. It can actually degrade the performance of the classifier \cite{Hastie01}. Thus, we consider the application of SIS as an important part of the proposed method.

\subsection{Simulated data}
The first simulated dataset has two classes of multivariate Gaussian distribution with the same covariance, i.e., $\bs{\Sigma}=\bs{I}_p$,  while the second simulated dataset is based on the nPCA model in \eqref{npca_model} with the number of nPCs, $r$, set to 5.  The only difference between the two classes lies in the centroids for the distributions.

\subsubsection{Simulation One}
There are two classes of $N(\bmu_1,\I_p)$ and $N(\bmu_2,\I_p)$ distributed independent variables of dimension $p=10000$ where $\bs{\mu}_1$ and $\bs{\mu}_2$ are all zeros, except that the first 100 components of $\bs{\mu}_2$ are equal to $0.5$. Here we are essentially generating data according the nPCA model \eqref{npca_model} with $r=0$. There are 100 training samples and 500 testing samples generated for each class.


The experiment is repeated 50 times and the values of the performance metrics CV err, TE and, \TE2, are given as the mean values of all the trials along with the standard deviation. We chose not to use SIS feature reduction in the simulations.

The results are shown in Table \ref{tab:sim1} where all the methods give relatively good test accuracies, except SVM which performs considerably worse than the other methods. The SCRDA and NSC methods perform very similarly while the \svn method, with $r=0$, has the lowest mean \TE2 value. The PLDA method performs worse than the other methods, with the exception of the linear SVM. The N/A for the SVM method in the table is because for the linear SVM we did not perform CV to determine the weight tuning parameter $C$, since its value turned out to be irrelevant.

\def\cmrw{1pt}
\begin{table}[htbp]
  \centering
  \caption{Simulation one with $\bf{\Sigma}=\I_p$.  The standard deviation is given in parenthesis}
    \begin{tabular}{rlll}
    \addlinespace
    \toprule
    Method & CV err & TE & \TE2 \\
    \midrule
    SVM & N/A & 217.9/1000 (12.1) & 217.9/1000 (12.13) \\
    SCRDA & 6.6/200 (2.8)  & 33.8/1000 (8.0) & 31.0/1000 (7.0) \\
    NSC & 6.9/200 (2.5) & 33.7/1000 (8.0) & 30.4/1000 (6.7) \\
	PLDA &  12.1/200 (3.8) & 49.0/1000 (9.8) & 43.5/1000 (11.2)\\
    \svn & 6.1/200 (2.7) & 34.5/1000 (10.5) & 29.6/1000 (7.2) \\
    \bottomrule
    \end{tabular}%
  \label{tab:sim1}%
\end{table}%

\subsubsection{Simulation Two}
This simulation is similar to the previous one with the exception that now the data is correlated according to the nPCA covariance matrix $\Om=\G\G^T+\s \bs{I}_p$. We set $r=5$ and use $p=10000$ features as before. Thus, sample $i$ from class $k$ is generated by using the nPCA model \eqref{npca_model}, $\x_i=\bmu_k+\G\bs{u}_i+\bs{\epsilon}_i$, where the mean vectors $\bmu_k$ are the same as in Simulation 1.  Only the first 100 rows of $\G$ are non-zero. Note that $\G$ is a $p \times r$ random matrix where we have chosen $r=5$. We use 50 trials while keeping $\G$ fixed. The number of training samples is 200 and the number of test samples is 1000. The results for this experiment are summarized in Table \ref{tab:sim2}.

The comparison methods perform poorly here, especially the SVM, PLDA and NSC methods but the proposed method performs very well, having a minimum test error \TE2 of an order of magnitude smaller than for the other methods.


\begin{table}[htbp]
  \centering
  \caption{Simulation two with data simulated using nPCA model \eqref{npca_model}.  The standard deviation is given in parenthesis}
    \begin{tabular}{rlll}
    \addlinespace
    \toprule
    Method & CV err & TE & \TE2 \\
    \midrule
    SVM & N/A & 434/1000 (12.9) & 434/1000 (12.9) \\
    SCRDA & 47.7/200 (6.0) & 222.1/1000 (46.6) & 212.2/1000 (46.2) \\
    NSC & 72.3/200 (5.90) & 390.16/1000 (28.8) & 370.68/1000 (22.5) \\
		PLDA & 80.2/1000 (8.3) & 427.1/1000 (18.3) & 414.0/1000 (14.2) \\
    \svn & 6.7/200 (2.92) & 22.4/1000 (8.36) & 19.8/1000 (4.8) \\
    \bottomrule
    \end{tabular}%
  \label{tab:sim2}%
\end{table}%

\subsection{Real Microarray Data.  The Golub dataset}
This cancer microarray dataset \cite{golub1999molecular} consists of 7129 gene expressions for 47 subjects having acute lymphoblastic leukemia (ALL) and 25 subjects who have acute myeloid leukemia (AML). The samples are divided into 38 training samples and 34 test samples, giving a total of 72 samples with 7192 features (genes).  The objective here is to correctly classify the samples to either the ALL group or the AML group. 

The results are given in Table \ref{tab:golub}. All the methods have a low misclassification rate for this dataset. The SVM method performs worst with 5 misclassified samples and the PLDA method is slightly better with 4 misclassifications, while the SCRDA and \svn methods give excellent results with \TE2=0 and TE=1.

\begin{table}[htbp]
  \centering
  \caption{Misclassification results for the Golub dataset. The parenthesis after \svn indicate what $r$ was used for white and gray matter, respectively}
    \begin{tabular}{rcccc}
    \addlinespace
    \toprule
    Method & CV err & Nonzeros & TE & \TE2 \\
    \midrule
    SVM & N/A & N/A & 5/34 & 5/34 \\
    SCRDA & 0/38 & 66 & 1/34 & 0/34 \\
    NSC & 1/38 & 3334 & 1/34 & 1/34 \\
		PLDA &2/38 & 1311 & 4/34 & 4/34\\
    \svn (2) & 1/38 & 404 & 1/34 & 0/34 \\
    \bottomrule
    \end{tabular}%
  \label{tab:golub}%
\end{table}%

\subsection{Neuropsychiatric CNVs Affecting Cognition}
In this experiment, gray and white matter segmented MRI data is used to classify subjects belonging to three groups, i.e., to classify a control group from two groups of subjects having neuropsychiatric CNVs that affect cognition. The control group (NoCNV) contains subjects with no large CNVs (NoCNV). The first group of subjects that have CNVs that affects cognition are subjects with 16p13.1 duplication. The second group of subjects with CNVs are subjects with 22q11.2 duplication.



The 16p13.1 duplication group contains subjects that have a DNA segment duplicated at chromosome 16 at location (locus) p13.11 and the 22q11.2 duplication CNV has a DNA segment duplicated at locus q11.2.  In \cite{Stefansson13} it is shown that controls carrying those CNVs perform considerably worse than controls having no CNV.



The groups used for the 16p13.1 and 22q11.2 duplication experiments are summarized in tables \ref{tab:AvsCsum} and \ref{tab:AvsD}, respectively. Both duplication groups have 18 subjects and they are age matched with the NoCNV group. We used two thirds of the data for training and one third for testing, so there are 24 subjects in the training set and 12 in the test set. We randomly assigned 6 males and 6 females out of the 18 subjects to each class in the training set.  
We used 5-fold CV to choose the optimal training parameters and we chose the folds such that the gender ratio was equal or very similar in each fold, i.e., stratified CV. The experiment was performed for both types of matter, white and gray. 

\begin{table}[htbp]
  \centering
  \caption{Summary of groups NoCNV and 16p13.1 dup. The groups have been matched for age}
    \begin{tabular}{rrrrrr}
    \addlinespace
    \toprule
    group & mean age & sd(age) & males & females\\
    \midrule
    NoCNV & 44.39 & 12.94 & 7 & 11 \\
    16p13.1 dup & 44.39 & 12.94 & 9 & 9 \\
    \bottomrule
    \end{tabular}%
  \label{tab:AvsCsum}%
\end{table}%

\begin{table}[htbp]
  \centering
  \caption{Summary of groups NoCNV and 22q11.2 dup. The groups have been matched for age}
    \begin{tabular}{rrrrrr}
    \addlinespace
    \toprule
    group & mean age & sd(age) & males & females  \\
    \midrule
    NoCNV & 42.67 & 13.11 & 7 & 11  \\
    22q11.2 dup & 42.67 & 13.11 & 10 & 8 \\
    \bottomrule
    \end{tabular}%
  \label{tab:AvsDsum}%
\end{table}%

\subsubsection{Results for 16p13.1 dup}
The results for the NoCNV vs. 16p13.1 dup classification experiment are shown in Table \ref{tab:AvsC}. The performance of the methods varies considerably for both types of matter. For white matter, the \svn and SCRDA methods perform best with zero misclassifications for the \TE2 metric. However, \svn performs better than SCRDA in terms of the TE metric. The SVM method performs worst in terms of \TE2 and the PLDA method performs second worst. 

Considering the gray matter results, we see that NSC, SCRDA, and the proposed method give the same results of $83.3\%$ accuracy for \TE2 while the SCRDA method has $50\%$ accuracy in TE. SVM performs worst with $75\%$ TE and \TE2 accuracy and PLDA is second worst with $66.7\%$ accuracy for both TE and \TE2. The \svn method turns out to have the best TE score. It seems that the SCRDA and NSC methods have trouble finding the optimal tuning parameters using CV.

Using SIS to reduce features approximately 100-fold down to 5000 is shown to be beneficial. For white matter, the \svn method has a perfect \TE2 of zero. For gray matter, the \svn method also benefits from the reduction of features, achieving perfect results, i.e., $100\%$ accuracy for both the TE metrics. Fig. \ref{fig:AvsC} shows the regions (voxels) of the brain which turned out to be the most discriminating between the NoCNV and 16p13.1 dup groups, according to the best \svn result for white matter in Table \ref{tab:AvsC}.

\subsubsection{Results for 22q11.2 dup}
For the NoCNV vs. 22q11.2 dup classification experiment, the results are summarized in Table \ref{tab:AvsD}. For white matter, the SVM and PLDA methods have TE and \TE2 values of 4 out of 12, which is $66.7\%$ classification accuracy. The \svn and both the SCRDA and NSC methods have a \TE2 value of 1, i.e., $91.7\%$ accuracy. Again, the \svn method has the lowest TE score, by a considerable margin.


For the gray matter, the \svn and NSC methods are the best performers considering both TE and \TE2. SIS feature selection failed to improve the \svn method, except that TE was improved for white matter. Once again, we see PLDA and SVM performing similarly.
Fig. \ref{fig:AvsD} shows the regions (voxels) of the brain which turned out to be the most discriminating between the NoCNV and 22q11.2 dup groups, according to the best \svn result for gray matter in Table \ref{tab:AvsD}.


In summary, for both  NoCNV vs. CNV duplication experiments, the proposed method clearly outperforms the other state-of-the-art comparison methods. The results obtained by the proposed method indicate that there are indeed significant structural differences in the brain between the NoCNV group and the CNV duplication groups. 

\begin{table*}[htbp]
  \centering
  \caption{NoCNV vs 16p13.1 dup misclassification results.  The parenthesis after \svn indicate what $r$ was used for white and gray matter, respectively}
    \begin{tabular}{rcccccccc}
    \addlinespace
    \toprule
     & \multicolumn{4}{c}{White matter} & \multicolumn{4}{c}{Gray matter} \\
    \cmidrule[\cmrw](lr){2-5} \cmidrule[\cmrw](lr){6-9}
    Method & CV err & Nonzeros & TE & \TE2 & CV err & Nonzeros & TE & \TE2 \\
		\cmidrule[\cmrw](lr){1-1} \cmidrule[\cmrw](lr){2-5} \cmidrule[\cmrw](lr){6-9}
    SVM & N/A & N/A & 4/12 & 4/12 & N/A & N/A & 3/12 & 3/12 \\
    SCRDA & 10/24 & 435 & 6/12 & 0/12 & 8/24 & 11439 & 6/12 & 2/12 \\
    NSC & 8/24 & 3 & 6/12 & 1/12 & 6/24 & 26 & 2/12 & 2/12 \\
		PLDA & 14/24 & 0 & 6/12 & 2/12 & 11/24 & 95024 & 4/12 & 4/12\\
    \svn(2,2) & 7/24 & 3584 & 1/12 & 0/12 & 7/24 & 230 & 2/12 & 2/12 \\
		\midrule
    \multicolumn{9}{c}{SIS (5000)} \\
		\midrule
    \svn(3,1) & 6/24 & 74 & 1/12 & 0/12 & 8/24 & 83 & 0/12 & 0/12 \\
    \bottomrule
    \end{tabular}%
  \label{tab:AvsC}%
\end{table*}%

\begin{table*}[htbp]
  \centering
  \caption{NoCNV vs 22q11.2 dup misclassification results. The parenthesis after \svn indicate what $r$ was used for white and gray matter, respectively}
    \begin{tabular}{rcccccccc}
    \addlinespace
    \toprule
     & \multicolumn{4}{c}{White matter} & \multicolumn{4}{c}{Gray matter} \\
    \cmidrule[\cmrw](lr){2-5} \cmidrule[\cmrw](lr){6-9}
    Methods & CV err & Nonzeros & TE & \TE2 & CV err & Nonzeros & TE & \TE2 \\
		\cmidrule[\cmrw](lr){1-1} \cmidrule[\cmrw](lr){2-5} \cmidrule[\cmrw](lr){6-9}
    SVM & N/A & N/A & 4/12 & 4/12 & N/A & N/A & 5/12 & 5/12 \\
    SCRDA & 9/24 & 9 & 4/12 & 1/12 & 9/24 & 349893 & 6/12 & 4/12 \\
    NSC & 7/24 & 145 & 4/12 & 1/12 & 9/24 & 7 & 5/12 & 3/12 \\
		PLDA & 11/24 & 52536 & 4/12 & 4/12 & 14/24 & 104758 & 5/12 & 5/12\\
    \svn(2,2) & 5/24 & 72 & 2/12 & 1/12 & 8/24 & 1960 & 4/12 & 2/12 \\
		\midrule
    \multicolumn{9}{c}{SIS (5000)} \\
		\midrule
    \svn(3,3) & 9/24 & 83 & 4/12 & 0/12 & 7/24 & 11 & 5/12 & 4/12 \\
    \bottomrule
    \end{tabular}%
  \label{tab:AvsD}%
\end{table*}%

\def\wi{0.9}
\begin{figure*}[htpb]
\begin{center}
\subfigure[Sagittal]{\includegraphics[width=\wi\linewidth]{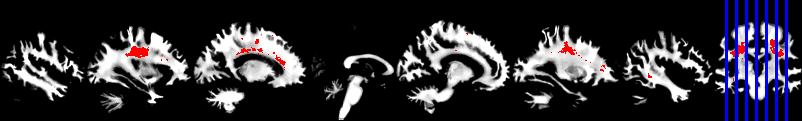}}\\
\subfigure[Coronal]{\includegraphics[width=\wi\linewidth]{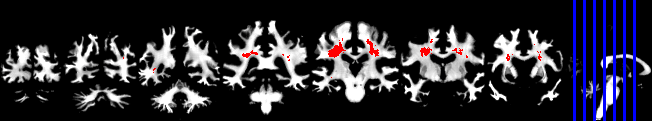}}\\
\subfigure[Axial]{\includegraphics[width=\wi\linewidth]{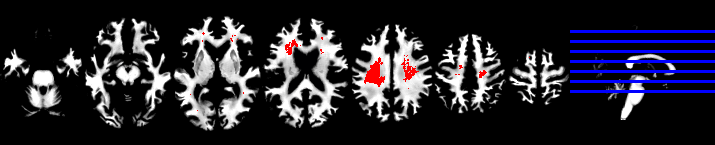}}
\caption{Discriminating voxels (red) for the \svn method for the NoCNV vs. 16p13.1 dup experiment (white matter).   The vertical and horizontal bars on the pictures to the right indicate what brain slices are plotted to the left.}
\label{fig:AvsC}
\end{center}
\end{figure*}

\def\wi{0.9}
\begin{figure*}[htpb]
\begin{center}
\subfigure[Sagittal]{\includegraphics[width=\wi\linewidth]{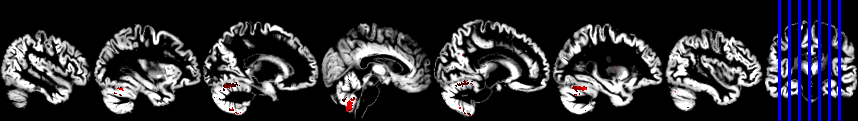}}\\
\subfigure[Coronal]{\includegraphics[width=\wi\linewidth]{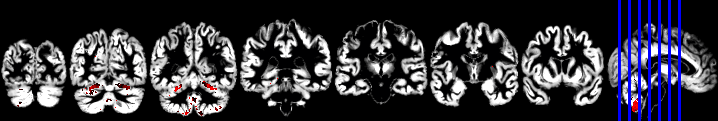}}\\
\subfigure[Axial]{\includegraphics[width=\wi\linewidth]{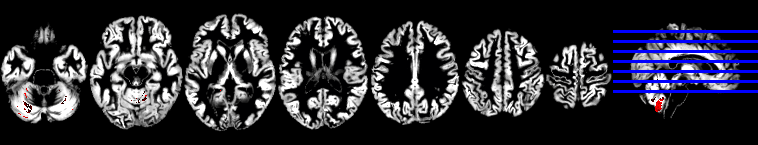}}
\caption{Discriminating voxels (red) for the \svn method for the NoCNV vs. 22q11.2 dup experiment (gray matter). The vertical and horizontal bars on the pictures to the right indicate what brain slices are plotted to the left.}
\label{fig:AvsD}
\end{center}
\end{figure*}




\subsection{Polygenic Risk Scores for Schizophrenia}
The aim of this final experiment is to determine if high or low PRS values for schizophrenia result in structural morphology of white or gray matter, that can be used for classification purposes.
PRS is defined as \cite{Dudbridge13}
\begin{eqnarray}
\hat{S}= \sum_{i=1}^m \hat{\beta}_i g_i. \nonumber
\end{eqnarray}
It is a linear combination of $m$ schizophrenia-associated alleles $g_i \in \{0,1,2\}$ weighted by effect size $\hat{\beta}_i$.  The number $m$ is selected based on some threshold on the $P$-value of the effect size, typically $P<0.1$.

For this study, we have structural MRI data for 54 control subjects with no large CNV, for which a PRS value based on the subject's genome has been calculated. There are two groups: low PRS, which consists of 27 subjects from the lower 5$\%$ tail of the PRS population distribution, and high PRS, that consists of 27 subjects from the upper 5$\%$ tail of the PRS population distribution. The odds ratio for schizophrenia in the low PRS group is 0.28, while the odds ratio for schizophrenia for the high PRS group is 4.63.  We use the same procedure as in the NoCNV vs. CNV duplication experiments to eliminate gender bias and we also split the data into training and testing datasets using the same proportions. Table \ref{tab:PRSsum} shows a summary of the two classes while the results for this experiment are summarized in Table \ref{tab:PRS}.




\begin{table}[htbp]
  \centering
  \caption{Summary of the low risk and high risk groups in PRS study}
    \begin{tabular}{rcccc}
    \addlinespace
    \toprule
     & \multicolumn{2}{c}{Low PRS} & \multicolumn{2}{c}{High PRS} \\
		\cmidrule[\cmrw](lr){2-3} \cmidrule[\cmrw](lr){4-5}
     & PRS & AGE & PRS & AGE \\
		\cmidrule[\cmrw](lr){1-1} \cmidrule[\cmrw](lr){2-3} \cmidrule[\cmrw](lr){4-5}
    \multicolumn{1}{l}{min} & -78.81 & 25 & -65.24 & 22 \\
    \multicolumn{1}{l}{max} & -74.41 & 63 & -63.00 & 66 \\
    \multicolumn{1}{l}{mean} & -75.50 & 47.48 & -64.30 & 47.85 \\
    \multicolumn{1}{l}{sd} & 1.02 & 10.91 & 0.68 & 11.01 \\
		\cmidrule[\cmrw](lr){1-1} \cmidrule[\cmrw](lr){2-3} \cmidrule[\cmrw](lr){4-5}
    \multicolumn{1}{l}{males} & \multicolumn{2}{c}{14} & \multicolumn{2}{c}{11} \\
    \multicolumn{1}{l}{females} & \multicolumn{2}{c}{13} & \multicolumn{2}{c}{16} \\
    \bottomrule
    \end{tabular}%
  \label{tab:PRSsum}%
\end{table}%

Considering the results in Table \ref{tab:PRS}, it is readily seen that these data are much more challenging than the NoCNV vs. duplication data. For white matter, all the comparison methods are performing at $50\%$ accuracy or even worse. The \svn method achieves an \TE2 score of 7 out of 18. However, it is quite obvious that it is difficult to obtain optimal tuning parameter values for \svn using CV for this dataset.

For gray matter, the results are somewhat similar. Again the comparison methods fail to find any discriminating features, while the \svn method has a \TE2 value of 5 out of 18, which translates to roughly $72\%$ accuracy, which is very acceptable, especially when given the poor performance of the comparison methods. As for the white matter, the high TE values are in stark contrast to the lower \TE2 values.  By using SIS to reduce the number of features down to 5000, we see that for white matter, the optimal misclassification rate goes down from 7 out of 18 to 5 out of 18, which is quite good. However, for the gray matter, SIS did not improve the results.


To summarize the findings of this experiment, it is clear that the comparison methods totally fail to discriminate between the two classes. The proposed method can do so, with up to $72\%$ accuracy, indicating that there are some detectable morphological differences in the brain between the low risk and high risk groups. 

\begin{table*}[htbp]
  \centering
  \caption{PRS study - 27 subjects from each tail of the PRS distribution. The parenthesis after \svn indicate what $r$ was used for white and gray matter, respectively}
    \begin{tabular}{rcccccccc}
    \addlinespace
    \toprule
     & \multicolumn{4}{c}{White matter} & \multicolumn{4}{c}{Gray matter} \\
    \cmidrule[\cmrw](lr){2-5} \cmidrule[\cmrw](lr){6-9}
    Methods  & CV err & Nonzeros & TE & \TE2 & CV err & Nonzeros & TE & \TE2 \\
		\cmidrule[\cmrw](lr){1-1} \cmidrule[\cmrw](lr){2-5} \cmidrule[\cmrw](lr){6-9}
    SVM & N/A & N/A & 9/18 & 9/18 & N/A & N/A & 9/18 & 9/18 \\
    SCRD & 18/36 & 0 & 9/18 & 9/18 & 13/36 & 323173 & 11/18 & 8/18 \\
    NSC & 14/36 & 0 & 10/18 & 10/18 & 14/36 & 15 & 9/18 & 9/18 \\
		PLDA & 17/36 & 109933 & 10/18 & 9/18 & 18/36 & 0 & 9/18 & 9/18\\
    \svn(1,1) & 15/36 & 196 & 11/18 & 7/18 & 16/36 & 179 & 11/18 & 5/18 \\
		\midrule
    \multicolumn{9}{c}{SIS (5000)} \\
		\midrule
    \svn(1,3) & 17/36 & 6 & 10/18 & 5/18 & 15/36 & 60 & 11/18 & 6/18 \\
    \bottomrule
    \end{tabular}%
  \label{tab:PRS}%
\end{table*}%

\section{Conclusions}
In this paper, we have considered the classification of big data where $p \gg n$.
In particular, we have focused on classification of 3-dimensional MRI images of the human brain in a genetic context. 

To address this issue we have proposed a  novel algorithm, based on LDA from the normal model viewpoint, where the estimation of the covariance of the observed data is performed by using the nPCA covariance model. Another important feature of the proposed method is its inherent ability to drop out variables that do not contribute to the class separation, using a vector $\ell_0$ penalty.  The proposed method depends on two tuning parameters, i.e., the number of noisy principal components and a sparsity tuning parameter, which were selected based on the cross-validation. We conducted a number of experiments using simulated data, real microarray data, and structural MRI data of both white and gray matter.

The experiment results indicate that when dealing with data of very high dimensionality with complex covariance structure such as the MRI data, our method compares well to other state-of-the-art big data methods.  An important item for future work is to develop methods, that are faster than cross-validation, for tuning parameter selection.

\section*{Appendix}
By expanding the quadratic of (\ref{EM}) and combining terms, we can write
\begin{eqnarray}
\mbox{EM}(\thet_0,\thet) &=& -\frac{p}{2} \log \s - \frac{\tr(\bs{S}_x)}{2 \s} - \frac{\tr(\G \B_0^T)}{\s} \nonumber \\
&-& \frac{\tr(\G \A_0 \G^T)}{2\s} - \frac{h}{2} \sum_{j=1}^p ||| \bs{a}_j|||_0, \nonumber
\end{eqnarray}
where
\begin{eqnarray}
\W_0 &=& \G_0^T \G_0 + \s_0 \bs{I}_r, \nonumber \\
\bs{u}_{i0} &=& \W_0^{-1} \G_0^T (\tilde{\x}_i - \bs{d}_{k0}), \quad i=1,...,r, \nonumber \\
\U &=& [\bs{u}_{i0}^T], \nonumber \\
\A_0 &=& \s_0 \W_0^{-1} + \frac{1}{n} \U^T \U, \nonumber \\
\bs{B}_0 &=& \frac{1}{n} \tilde{\X}^T \U, \nonumber \\
\bs{S}_{x}&=&\frac{1}{n} \sum_{k=1}^K \sum_{i \in C_k} (\tilde{\x}_i-\bs{d}_{k})(\tilde{\x}_i-\bs{d}_{k})^T. \nonumber
\end{eqnarray}
Maximization w.r.t. $\G$ is equivalent to minimization of the following cost function w.r.t. $\G$
\begin{eqnarray}
J_1 &=& \sum_{j=1}^p J(\bs{g}_j) \nonumber \\
&=& \sum_{j=1}^p \left( \frac{1}{2} \bs{g}_j^T \A_0 \bs{g}_j - \bs{g}_j^T \bs{b}_{j0} + \frac{h \s}{2} ||| \bs{a}_j |||_0 \right). \nonumber
\end{eqnarray}
The cost function is separable in the rows of $\G$, so we optimize it for each $\bs{g}_j$ individually.  Due to the $l_0$ penalty, the cost function is not differentiable at zero, so we proceed in two steps.  First we assume that $\bs{g}_j \neq \bs{0}$ and find the optimal solution, then we compare the resulting cost to the cost when $\bs{g}_j = \bs{0}$.  Assuming that $\bs{g}_j \neq \bs{0}$ and differentiating and setting $J$ equal to zero yields
\begin{eqnarray}
\bs{g}_{j} = \A_0^{-1} \bs{b}_{j0}. \nonumber
\end{eqnarray}
A comparison with the $\bs{g}_j = \bs{0}$ solution yields
\begin{eqnarray}
J(\bs{g}_j) - J(\bs{0}) = - \tau_j^2 + h \s \geq 0, \nonumber
\end{eqnarray}
where
\begin{eqnarray}
\tau_j^2 = \bs{b}_{j0}^T \A_0^{-1} \bs{b}_{j0} + \sum_{k=1}^K \frac{n_k}{n}(\hat{\mu}_{kj}-\hat{\mu}_j)^2.
\end{eqnarray}
The $\bs{g}_j=\bs{0}$ solution is picked if $h \s_0 > \tau_j^2 \s$ and the minimizer is given by
\begin{eqnarray}
\bs{g}_{j1} &=& \bs{A}_0^{-1} \bs{b}_{j0} I(\tau_j^2 \geq h \s_{0} ), \quad j=1,...,p. \nonumber
\end{eqnarray}
By using the same optimization method, we can optimize the EM functional w.r.t. $d_{kj}$ (note that it depends on the EM functional through $\bs{S}_x$).  The optimization yields
\begin{eqnarray}
d_{kj1}&=& (\hat{\mu}_{kj}-\hat{\mu}_j)I(\tau_j^2 \geq h \s_{0}), \quad j=1,...,p, k=1,...,K. \nonumber
\end{eqnarray}
The optimization of the EM functional w.r.t. $\s$ easily yields
\begin{eqnarray}
\s_1 = \frac{1}{p}[\tr(\bs{S}_x)-2\tr(\B_0^T \G_1) + \tr(\A_0 \G_1^T \G_1)], \nonumber
\end{eqnarray}
which can be simplified to
\begin{eqnarray}
\sigma^2_1= \frac{1}{p} \sum_{j  \in \mathcal{I}} \left( S_{xjj}- \bs{b}_{j0}^T\A_0^{-1} \bs{b}_{j0} \right) + \frac{1}{p}\sum_{j \in \mathcal{I}^c} S_{xjj}. \nonumber
\end{eqnarray}








\bibliographystyle{IEEEtran}

\bibliography{IEEEabrv,thebib}
\begin{IEEEbiography}[{\includegraphics[width=1in,height=1.25in,clip,keepaspectratio]{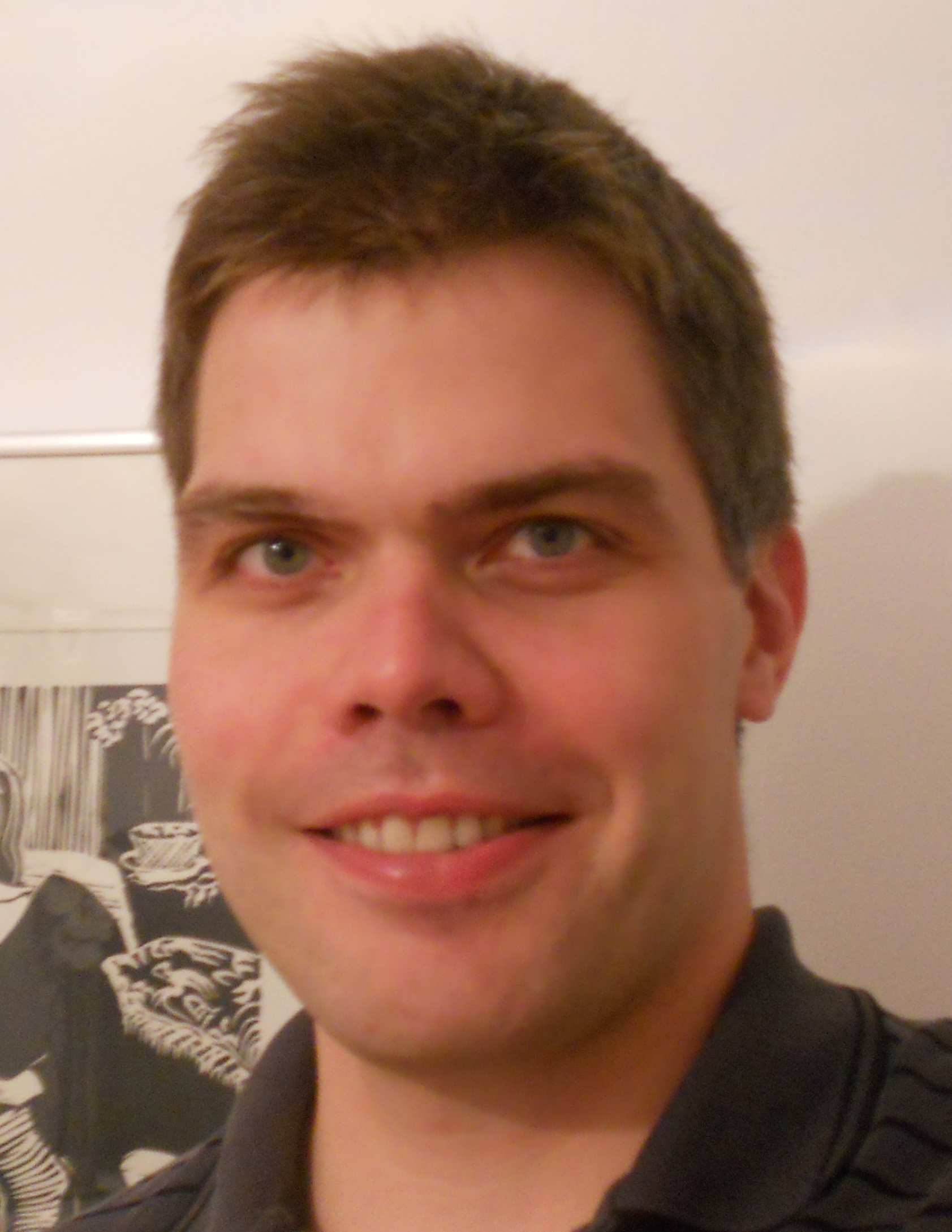}}]{Magnus Orn Ulfarsson}
Magnus O. Ulfarsson received the B.S. and the M.S. degrees from the University of Iceland in 2002, and the Ph.D. degree from the University of Michigan in 2007.   He joined University of Iceland in 2007 where he is currently a professor.  He has been affiliated with Decode Genetics, Reykjavik, Iceland since 2013.  His research interest include statistical signal processing, genomics, medical imaging, and remote sensing. \end{IEEEbiography}

\begin{IEEEbiography}[{\includegraphics[width=1in,height=1.25in,clip,keepaspectratio]{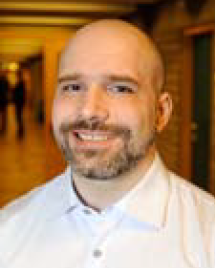}}]{Frosti Palsson}
Frosti Palsson received the B.S. and the M.S. degrees in electrical engineering from the University of Iceland in 2012 and 2013. He is currently working towards his Ph.D. degree at the University of Iceland. His research interests include image fusion in remote sensing, image and signal processing.\end{IEEEbiography}

\begin{IEEEbiography}[{\includegraphics[width=1in,height=1.25in,clip,keepaspectratio]{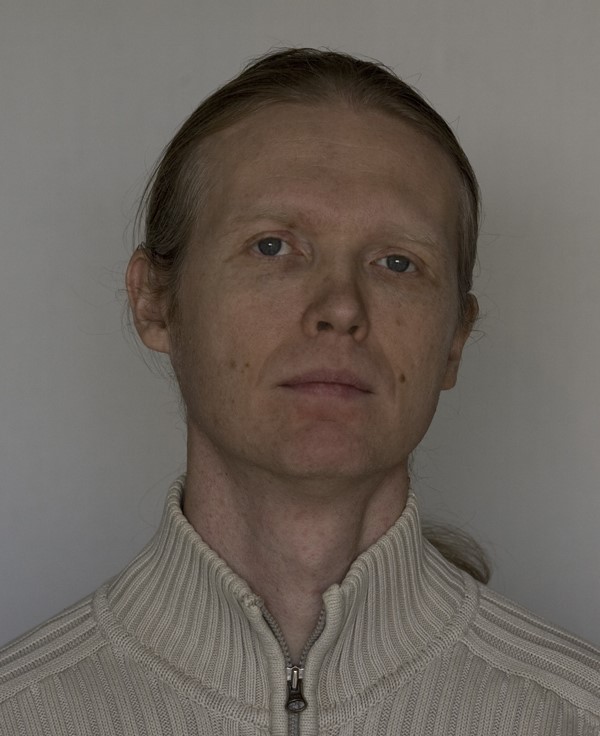}}]{Jakob Sigurdsson}
 (S'10) received the B.S. and M.S. degree in 2011 and the Ph.D. degree in 2015, from the University of Iceland. His research interests include statistical signal processing, remote sensing and image processing.
\end{IEEEbiography}

\begin{IEEEbiography}[{\includegraphics[width=1in,height=1.25in,clip,keepaspectratio]{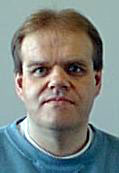}}]{Johannes R. Sveinsson}
Johannes R. Sveinsson received the B.S. degree from the University of Iceland, Reykjavík, and the M.S. and Ph.D. degrees from Queen's University, Kingston, ON, Canada, all in electrical engineering. He is currently Professor with the Department of Electrical and Computer Engineering, University of Iceland, he was with the Laboratory of Information Technology and Signal Processing from 1981 to 1982 and, from November 1991 to 1998, the Engineering Research Institute and the Department of Electrical and Computer Engineering as a Senior Member of research staff and a Lecturer, respectively. He was a Visiting Research Student with the Imperial College of Science and Technology, London, U.K., from 1985 to 1986. At Queen's University, he held teaching and research assistantships. His current research interests are in systems and signal theory. Dr. Sveinsson received the Queen's Graduate Awards from Queen's University. He is a co-recipient of the 2013 IEEE GRSS Highest Impact Paper Award.
\end{IEEEbiography}

\end{document}